\documentclass[%
superscriptaddress,
twocolumn,
amsmath,
amssymb,
aps,prl
]{revtex4-1}

\usepackage{graphicx}  
\usepackage{dcolumn}  
\usepackage{bm}            
\usepackage{amsmath}
\usepackage{xfrac}
\usepackage{xcolor}
\usepackage[colorlinks=true, citecolor=red]{hyperref}   
\usepackage{hhline}
\usepackage{diagbox}
\usepackage{siunitx}
\usepackage{enumitem}
\usepackage[normalem]{ulem}

\newcommand\redout{\bgroup\markoverwith
	{\textcolor{red}{\rule[.5ex]{2pt}{1pt}}}\ULon}

\newcommand{\meV}{{\textrm{ meV}}}
\newcommand{\colorrev}{\color{black}}
\begin{document}
	
	\title{Nematic excitonic insulator  in transition metal dichalcogenide moir\'e heterobilayers}
	\author{Ming Xie}
     		\email{mingxie@umd.edu}
		\affiliation{Condensed Matter Theory Center and Joint Quantum Institute, Department of Physics, 
			                University of Maryland, College Park, Maryland 20742, USA}
		
	\author{Haining Pan}
		\affiliation{Condensed Matter Theory Center and Joint Quantum Institute, Department of Physics, 
			                University of Maryland, College Park, Maryland 20742, USA}
		
	\author{Fengcheng Wu}
		\affiliation{School of Physics and Technology, Wuhan University, Wuhan 430072, China}
		\affiliation{Wuhan Institute of Quantum Technology, Wuhan 430206, China}
		
	\author{Sankar Das Sarma}
    	\affiliation{Condensed Matter Theory Center and Joint Quantum Institute, Department of Physics, 
    		                University of Maryland, College Park, Maryland 20742, USA}
    	                
	\date{\today}
	
	\begin{abstract}
		
		We study the effect of inter-electron Coulomb interactions on the displacement field induced topological phase transition in
		transition metal dichalcogenide (TMD) moir\'e heterobilayers. 
		We find a nematic excitonic insulator (NEI) phase that breaks the moir\'e superlattice's three-fold rotational symmetry
 		and preempts the topological phase transition in both AA and AB stacked heterobilayers when the interlayer tunneling is weak,
 		or when the  Coulomb interaction is not strongly screened.
		The nematicity originates from the frustration between the nontrivial spatial structure of the interlayer tunneling,
		which is crucial to the existence of the topological Chern band, and the interlayer coherence induced by the Coulomb interaction
		that favors uniformity in layer pseudo-spin orientations.
		We construct a unified effective two-band model that captures the physics near the band inversion and applies to both AA and AB stacked
		heterobilayers. Within the two-band model the competition between the NEI phase and the Chern insulator phase can be understood as 
		the switching of the energetic order between the $s$-wave and the $p$-wave excitons upon increasing the interlayer tunneling.
		
	\end{abstract}
	
	\maketitle

\emph{Introduction}.---\noindent
The field of moir\'e twistronics has been making tremendous strides
ever since the discovery of correlated insulating states and superconductivity simultaneously
in magic angle twisted bilayer graphene a few years ago \cite{CaoInsulator, CaoSuper}.
Twisted moir\'e systems have become a testbed for topology, strong correlation and superconductivity, 
and the rich interplay among them 
\cite{Efetov, YoungDean, Gordon, YoungQAHE, MakHubbard2020, Dean2020, WignerCrystal, Shan2020, LeRoy,
StripePhase,  MakMott2021, QuantumCriticality, Pasupathy2021, Cui2021}.
Until a year ago, topological phenomena were only observed in graphene-based moir\'e materials \cite{Gordon, YoungQAHE}.
The recent experimental observation \cite{MakQAHE2021} of quantum anomalous Hall (QAH) effect in half-filled ($\nu=1$) AB-stacked 
MoTe$_2$/WSe$_2$ moir\'e heterobilayers demonstrated the existence
of topological bands in TMD based moir\'e bilayers.

Topological bands in TMD moir\'e superlattices were first
predicted \cite{WumoireTI} in AA-stacked homobilayers with a small interlayer twist angle \cite{PanMIT, PanMoireHubbard, MillisHF, FuMagicAngle}.
There it was understood that the interlayer tunneling with a
real space skyrmion pattern is crucial (albeit not sufficient) for having a non-zero Chern number \cite{WumoireTI, Pan2020}.
In TMD heterobilayers,
interlayer tunneling is often treated as a negligible perturbation because of 
the large bandoffset between the valence bands of  the two layers,
and holes are thus confined to only the low energy layer \cite{WumoireHubbard}.
Many interesting strongly correlated phases, such as magnetic, Wigner crystal,
and Mott insulating phases, are being actively investigated in these systems
\cite{ MakHubbard2020, WignerCrystal, Shan2020, ShanMIT,
	StripePhase,  MakMott2021, Pasupathy2021, Cui2021, AllanHF, AllanED}.
The QAH experiment \cite{MakQAHE2021} suggested the origin of topology 
as arising from the field-induced band inversion \cite{FuPNAS2021},
but how (and whether) the strongly correlated nature of the moir\'e bands enters the topological physics
remains unclear \cite{FuInvert, LawStrain, PanHF}.

In this Letter, we address this key issue by considering realistic Coulomb interactions screened
by nearby metallic gates.
We perform un-projected self-consistent mean-field calculations to
continuously track how the interacting ground state evolves across the topological band inversion.
At filling factor $\nu=1$, we find a nematic excitonic insulator (NEI) phase,
which is present in the phase diagram of both AA and AB stacked heterobilayers,
and has a universal mechanism, that is, 
{\colorrev the layer pseudo-spin winding required by nontrivial topology
is always in frustration with uniformity of the spontaneous interlayer coherence favored by the Coulomb interaction.}
The NEI phase is a compromise between these two trends, 
and preempts the phase transition to the Chern insulator (CI) state at weak interlayer tunneling.
We develop an effective two-band model which illustrates
the physics of this competition as the switching of the energetic order
between the $s$-wave and the $p$-wave exciton condensates.
Our work provides a physically appealing basis for the observed topology.

{\colorrev
In TMD moir\'e heterobilayers at $\nu=1$, 
valley polarized and intervalley coherent orders
lead to insulating states of a drastically different nature:
the former with a gap at field-induced band inversion 
while the latter with a valley hybridization gap.
As a consequence, the valley coherent insulating phases (e.g. for AB-stacked MoTe$_2$/WSe$_2$ discussed in \cite{PanHF})
are largely agnostic of the frustration between interaction and topology
occurring dominantly near the band inversion, which is the focus of the present study.}

\emph{Moir\'e Hamiltonian}.---\noindent
We start by introducing a general continuum model Hamiltonian for moir\'e heterobilayers
when the lattice mismatch and the interlayer twist angle are small \cite{WumoireTI,FuPNAS2021}.
It builds on the effective $\bm{k} \cdot \bm{p}$ model for the highest spin-split valence bands
near the $+K$ and $-K$ valleys of the isolated monolayers \cite{Di2012}.
The Hamiltonian for the valley $+K$ takes the form
\begin{align}
	\mathcal{H}_s^{+} = 
	\begin{pmatrix}
		H^0_{t}(\bm{k}) + \Delta(\bm{r}) & T_s(\bm{r})\\
		T_s^\dagger(\bm{r})  & H^0_{b}(\bm{k})- \Delta E_v
	\end{pmatrix},
	\label{moireHam}
\end{align}
where $s=0$ for AA and $s=1$ for AB stacked bilayers, respectively.
The physical origin of $s$ will be discussed shortly.
$H^0_{t(b)}(\bm{k})=-\hbar^2(\bm{k}-\bm{\kappa}(\bm{\kappa}'))^2/2m_{t(b)}$ is the effective Hamiltonian 
for the top (bottom) layer with effective mass $m_{t(b)}$. 
$ \Delta E_v$ is the band offset between the two layers tunable by displacement field.
$\Delta(\bm{r})$ is the intralayer moir\'e potential \cite{WumoireHubbard} for the top layer which can be expanded as
\begin{align}
	\Delta(\bm{r})= 2V\sum_{i=1,3,5} \cos(\bm{g}_{i}\cdot\bm{r}+\phi),
\end{align}
where $V$ and $\phi$ are the amplitude and the phase of moir\'e potential.
$\bm{g}_1={4\pi}/{\sqrt{3}a_M}(1, 0)$ and $\bm{g}_i=(\hat{\mathcal{R}}_{\pi/3})^{i-1}\bm{g}_1$ 
are moir\'e reciprocal lattice vectors 
where $\hat{\mathcal{R}}_{\pi/3}$ is counter-clockwise rotation around $z$ axis by $\pi/3$.
Here we neglect the intralayer moir\'e potential for the bottom layer 
since it has a negligible effect on states near the band inversion.

The interlayer tunneling is given by
\begin{align}
	T_s(\bm{r}) = t(1 + \omega^{s} e^{i\bm{g}_2\cdot\bm{r}} + \omega^{2s} e^{i\bm{g}_3\cdot\bm{r}}),
\end{align}
where $t$ is the tunneling strength and $\omega=e^{i2\pi/3}$. 
$s$ is determined by the difference in angular momentum $j$ between
the top and bottom layer states at $+K$, $s=j_t-j_b$ \cite{FuPNAS2021}.
Here $j$ is defined under three-fold rotation and includes both spin and orbital angular momentum 
contributions from the isolated monolayers \cite{Di2012}.
For AA-stacked bilayers, $j_t=j_b=1/2$ whereas
for AB-stacked bilayers, $j_t=-j_b=1/2$.
Microscopically, the anti-alignment of the spin in AB-stacked bilayers
strongly suppresses the interlayer tunneling.

\begin{figure}[t!]
	\centering
	\includegraphics[width=0.49\textwidth]{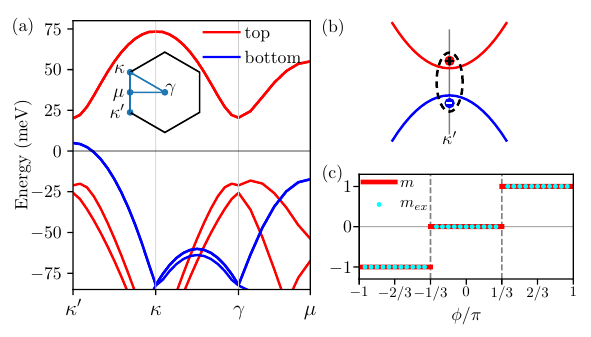}%
	\caption{Mean-field ground states in the zero interlayer tunneling limit. 
		(a) Hartree-Fock energy dispersion for AB-stacked MoTe$_2$/WSe$_2$
		with $t=0$, $(V, \phi)=(4.1\meV, 14^{\circ}),\ \Delta E_v=10\meV$, and $\epsilon=10$.
		Inset plots the moir\'e Brillouin zone. 
		(b) Schematic plot of the interlayer exciton formed near the band-edge at $\bm{\kappa}'$.
		(c) Moir\'e angular momentum of the top layer electron $m$ and of the exciton condensate $m_{ex}$ as a function of phase angle $\phi$.
	}
	\label{fig:tzero}
\end{figure}

\emph{Topological band inversion}.---\noindent
The key ingredient underlying a topological band inversion is that
the states near the two opposing band edges  being inverted have different orbital characters.
In two-dimensional systems, this is indicated by having different crystal angular momentum, 
i.e., different eigenvalues under crystal rotational symmetry operation \cite{Fang2012, FangPGS, Hughes2011, Kane2018}.
The general moir\'e Hamiltonian in Eq.~(\ref{moireHam}), however, does not always guarantee
that the displacement field-induced band inversion is topological.
In addition to the angular momentum $j_t$ inherited from the isolated monolayer,
the top layer state at the band edge $\bm{\kappa}'$ (for valley $K$) acquires an additional angular momentum $m$,
due to the presence of the intralayer moir\'e potential $V(\bm{r})$.
The total angular momentum of the top layer state at $\bm{\kappa}'$ 
is then $j_t+m$,
and the condition for the band inversion to be topological becomes
\begin{align}
	j_t+m\neq j_b. \label{topocond}
\end{align}
The value of $m$ is determined by the moir\'e potential phase angle $\phi$ in a piecewise manner
as shown in Fig.~\ref{fig:tzero}(c).
When $\phi=\pm\pi/3$ and $\pm\pi$,
$m$ is not well defined due to band degeneracy, 
which we exclude in this study.
The angular momenta on both sides of Eq.~(\ref{topocond}) are defined modulo 3.
For example, $m=0$ in AB-stacked MoTe$_2$/WSe$_2$ with $\phi=14^{\circ}$ \cite{FuPNAS2021}
and $m=-1$ in AA-stacked WSe$_2$/MoSe$_2$ with $\phi=-94^{\circ}$ \cite{WumoireTI}, 
both satisfying the topological condition [Eq.~(\ref{topocond})].

\emph{Mean-field theory}.---\noindent
The goal of this Letter is to study the effect of Coulomb
interaction on the topological phase transition.
We adopt un-projected self-consistent Hartree-Fock mean-field theory 
which provides a good description for insulators (see Supplemental Material for details \cite{SM}).
We focus on moir\'e band filling factor $\nu=1$
at which the QAH effect was observed \cite{MakQAHE2021}.

At filling factor $\nu=1$, there are two ways the ground state
can lower its energy: one is by gaining exchange energy by
fully occupying (with holes) one of the two valleys,
and the other is by developing spontaneous coherence between the two valleys,
which leads to valley density wave states breaking the moir\'e 
translational symmetry \cite{PanHF}.
We find that the valley polarized phases are in general favored by
stronger Coulomb interaction over the intervalley coherent (IVC) phases.
This is consistent with the findings in heterobilayers in the absence of
the displacement field \cite{AllanHF, AllanED, withfield}.
On the other hand, establishing interlayer coherence strongly affects the
valley polarized phases
and lowers their total energies.
The valley polarized phases dominate the phase diagram
at more realistic values of the interaction strength, $\epsilon^{-1}\sim0.1$.
For these reasons we focus on valley polarized phases throughout this work.

\begin{figure}[t!]
	\centering
	\includegraphics[width=0.49\textwidth]{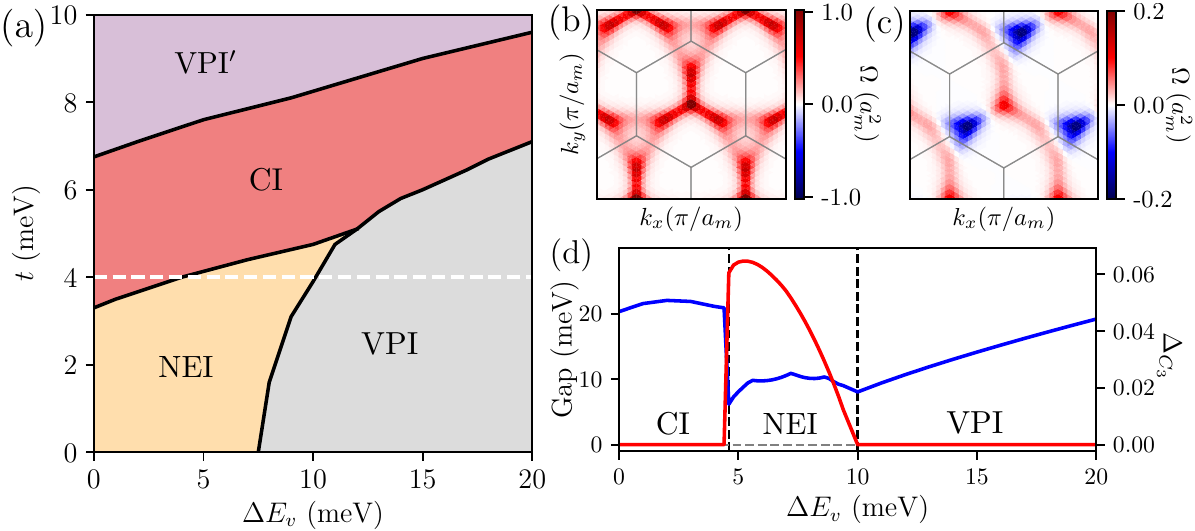}%
	\caption{(a) Mean-field phase diagram of AB-stacked MoTe$_2$/WSe$_2$ heterobilayer 
		at $\nu=1$ with $\epsilon=11$. 
		Berry curvature of the highest moir\'e band in the (b) Chern insulator (CI) phase 
		with $t=4\meV,\ D=1\meV$
		and (c) nematic excitonic insulator (NEI) phase with $t=1\meV,\ D=1\meV$.
		(d) Nematic order $\Delta_{C_3}$ and the direct energy gap along the white dashed line
		in (a) with $t=4.0\meV$. VPI and VPI$'$ are trivial valley polarized insulating phases.} 
	\label{fig:phaseAB}
\end{figure}

\emph{Exciton insulator in the zero interlayer tunneling limit}.---\noindent
We start by looking at the mean-field ground states when $t=0$.
At large $\Delta E_v$, the valley polarized ground state is also layer polarized with all
holes residing in the top layer.
Decreasing $\Delta E_v$ reduces the bandgap $E^{tb}_g$ between the highest moir\'e bands in the top and the bottom layers.
Fig.~\ref{fig:tzero}(a) plots the quasiparticle bandstructure when $E^{tb}_g$ 
is slightly larger than the exciton binding energy $E_b$. 

When $E^{tb}_g<E_b$, the exciton condensate state becomes the ground state
and establishes spontaneous interlayer coherence.
For ground states polarized
in valley $+K$, the mean-field Fock exchange self-energy takes the form
\begin{align}
	\Sigma^{+}_{F}(\bm{k}) &=
	\begin{pmatrix}
        \Delta_{t}(\bm{k}) & \Delta_{tb}(\bm{k})\\
        (\Delta_{tb})^{\dagger}(\bm{k})& \Delta_{b}(\bm{k})
	\end{pmatrix},\ \ \ 
\Sigma^{-}_{F}(\bm{k})=0
\end{align}
where the superscript $\pm$ stands for $\pm K$ valley and the subscript $t(b)$ stands for the top (bottom) layer.
Each element by itself is a matrix in the moir\'e reciprocal lattice vector basis with the form
$\Delta_{\bm{G}\bm{G}'}(\bm{k})$.
The interlayer Fock self-energy $\Delta_{tb}(\bm{k})$ can be defined as the condensate order parameter, the onset of which signifies the onset of the exciton condensation.
For AB-stacked MoTe$_2$/WSe$_2$, we find $E_b=12\meV$.

The condensate order parameter transforms under three-fold rotation $C_3$ around $z$ axis as
\begin{align}
	U_{\hat{C}_3}  \Delta_{tb}(\hat{R}_3\bm{k})(U_{\hat{C}_3})^{-1} 
	=\omega^{m_{\rm ex}}\Delta_{tb}(\bm{k}),
	\label{deltatransform}
\end{align}
where $[U_{\hat{C}_3}]_{\bm{G}\bm{G'}}=\langle \bm{G}|R_{3}\bm{G}'\rangle$ is the matrix form of the $\hat{C}_3$ operator.
$m_{ex}$ is an integer defined modulo 3
and can be viewed as the angular momentum of the condensing excitons (factoring out $j_t-j_b$).
$m_{ex}$ depends piece-wise on $\phi$
as shown in Fig.~\ref{fig:tzero}(c), and $m_{ex}=m$ inside all three regions (except at the special values $\phi=\pm \pi/3,\pm\pi$).
Intuitively, $m_{ex}$ has contributions from the moir\'e angular momentum of the electron and the hole individually, and from the relative orbiting motion between the two.
The fact that $m_{ex}$ is always equal to $m$, the electron angular momentum,
suggests that the relative orbiting motion is always $s$-wave like with zero angular momentum,
independent of $\phi$. (The hole angular momentum is zero).
This is consistent with  $s$-wave exciton condensation.

\begin{figure}[t!]
	\centering
	\includegraphics[width=0.49\textwidth]{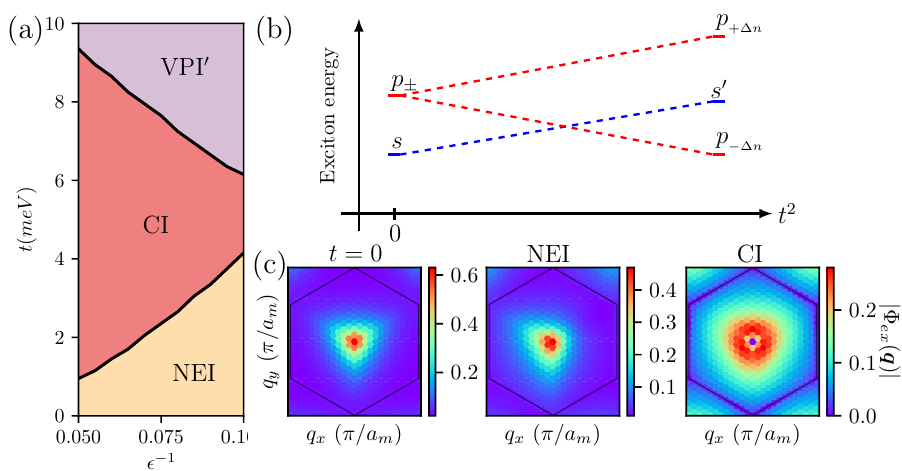}%
	\caption{(a) Dependence of the mean-field ground states on interaction strength.
		The band offset is fixed at $\Delta E_v=0$.
	(b) Schematic plot of exciton energies as a function of $t$. 
	$\Delta n=\pm 1$.
     (c) Projected exciton wavefunction amplitude $|\Phi_{ex}(\bm{q})|$ for 
     the $t=0$ case with $D=6$\meV,  the NEI phase with $t=3\meV, D=8\meV$, and the CI phase with $t=6\meV, D=14\meV$.}
	\label{fig:exciton}
\end{figure}

\emph{Phase diagram}.---\noindent
We now consider the mean-field ground states in the presence of finite interlayer tunneling.
{\colorrev Unlike the Coulomb interaction, the interlayer tunneling generates winding of the layer pseudospin
required for nontrivial topology.}
This implies a competition between the two and
motivates us to study the ground states from weak to strong interlayer tunneling limits.

The phase diagram in Fig.~\ref{fig:phaseAB} is calculated, as an example,
using the parameters of AB-stacked MoTe$_2$/WSe$_2$ bilayer
$(V, \phi)=(4.1meV, 14^{\circ})$, with $t$ being a varying parameter.
The most prominent feature of the phase diagram
is the nematic exciton insulator (NEI) phase at small $t$,
which spontaneously breaks the $C_3$ rotational symmetry.
The interlayer tunneling term transforms under $C_3$ as
\begin{align}
[U_{\hat{C}_3}T(U_{\hat{C}_3})^{-1}]_{\bm{G}\bm{G}'}= \omega^{-s}T_{\bm{G}\bm{G}'}
\end{align}
where $T_{\bm{G}\bm{G}'}$ is the Fourier transform of $T(\bm{r})$.
From the topological condition in Eq.~(\ref{topocond}), it immediately follows that $\omega^{-s}\neq \omega^{m_{ex}}$ 
(because $s=j_t-j_b$ and $m_{ex}=m$). 
This means that the interlayer coherence induced by interaction 
and that induced by interlayer tunneling transform differently and are therefore \textit{frustrated} in momentum space.
When $t$ increases gradually from zero,
the exciton insulator ground state has to adjust to the frustration by breaking the $C_3$ symmetry
as our calculations demonstrate.

The ground state, depending on the strength of $t$, transitions from the valley polarized trivial insulator (VPI) phase at large $\Delta E_v$,
to either the NEI phase or the Chern insulator (CI) phase (with Chern number $C=1$) as $\Delta E_v$ is reduced.
The CI state is favored by stronger interlayer tunneling,
 and its exchange self-energy $\Delta_{tb}(\bm{k})$ transforms in the same way as $T$,
restoring the $C_3$ symmetry.
Figures~\ref{fig:phaseAB}(b) and (c) show the Berry curvature of the highest moir\'e band in CI and NEI states, respectively;
{\colorrev the Chern number of the NEI state is zero.}
At even larger $t$, the bandgap closing and reopening at $\bm{\gamma}$ point occur, 
similar to the non-interacting band behavior \cite{gammatouching},
leading to another trivial state VPI$'$,
which is not adiabatically connected to the VPI state.
{\colorrev The phase diagram in Fig. 2(a) does not contain intervalley coherent phases
predicted in the earlier work \cite{PanHF} at large interlayer tunneling $t=12 meV$ and
favored by weaker interaction strength as discussed above.}
The stronger interaction strength and small critical value of $t$ at the NEI-CI transition in Fig.~\ref{fig:phaseAB}(a) are closer to realistic parameters according to ab initio calculation \cite{FuPNAS2021} and capacitance measurement \cite{MakQAHE2021}, supporting the relevance of the predicted NEI-CI competition.

We define an order parameter for nematicity 
${\Delta}_{C_3}= A^{-1}\sum_{\bm{k}}1 -|\langle\psi_{1, \mathcal{R}_3\bm{k}}|C_3|\psi_{1,\bm{k}}\rangle|$
in terms of the 
wavefunction $|\psi_{+1,\bm{k}}\rangle$ of the highest band in valley $+K$.
Fig.~\ref{fig:phaseAB}(d) plots ${\Delta}_{C_3}$ as a function of $\Delta E_v$ at fixed $t=4meV$.
At the VPI/NEI transition, ${\Delta}_{C_3}$ grows continuously from zero, 
suggesting a continuous phase transition,
while at the NEI/CI transition ${\Delta}_{C_3}$ drops to zero abruptly, characteristic of a first order transition;
the direct energy gap [Fig.~\ref{fig:phaseAB}(d)] shows the same behavior across the two phase transitions.

When the interaction strength, characterized by $\epsilon^{-1}$, is tuned, we find that the phase diagram remains qualitatively the same.
Figure~\ref{fig:exciton}(a) plots the phase diagram as a function of $\epsilon^{-1}$ at fixed $\Delta E_v=0$.
The expansion of the NEI region as $\epsilon^{-1}$ increases is consistent with Coulomb interactions favoring
the NEI state.

Because of the frustration, the NEI state can be viewed as a ``distorted" $s$-wave exciton condensate.
This can be seen by examining the exciton wavefunction $\Phi_{ex}(\bm{q})$,
when the exciton density is low (see details in the SM \cite{SM}).
$\bm{q}$ is the momentum measured from $\bm{\kappa}'$.
Figure~\ref{fig:exciton} (c) shows the amplitude of $\Phi_{ex}(\bm{q})$ which illustrates the nematic nature of the $s$-wave excitons at finite $t$.
The CI state, on the other hand, has the character of a $p$-wave condensate
with a node at $\bm{q}=0$.

\begin{figure}[t!]
	\centering
	\includegraphics[width=0.49\textwidth]{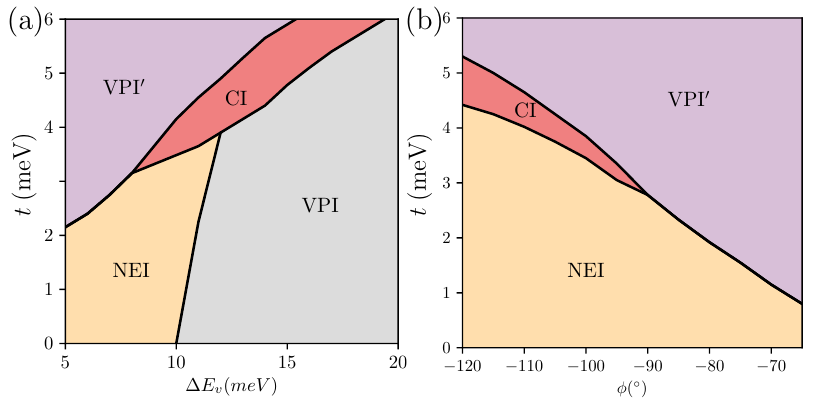}%
	\caption{(a)Mean-field phase diagram for AA-stacked WSe$_2$/MoSe$_2$ heterobilayer at $\nu=1$ for $\epsilon=11$.
		(b) The dependence of the ground state on the phase angle $\phi$ 
	     for fixed $\Delta E_v=8.0\meV$. }
	
	\label{fig:phaseAA}
\end{figure}

The physics we discussed above is independent of stackings and 
applies to AA-stacked heterobilayers as well when the topological condition Eq.~(\ref{topocond}) is satisfied.
To verify this, we perform self-consistent mean-field calculations for AA-stacked WSe$_2$/MoSe$_2$ bilayer with $s=0$
and $(V,\phi)=(6.6\meV,-94^{\circ})$.
The resulting phase diagram, shown in Fig.~\ref{fig:phaseAA}(a),
demonstrates the presence of the NEI phase.
The quantitative difference compared to the AB-stacked MoTe$_2$/WSe$_2$
is attributed to the difference in the continuum model parameters.
The phase diagram has a strong dependence on $\phi$, as shown in Fig.~\ref{fig:phaseAA}(b),
mainly because the effective electron mass near $\bm{\kappa}'$,
and thus the exciton binding energy $E_b$,
depend sensitively on $\phi$.

\emph{Effective two-band model}.---\noindent
Guided by our findings of the same physics 
in both AA- and AB-stacked heterobilayers,
we construct an effective theory that unifies the physics
near the topological band inversion in both cases.
The simplest effective Hamiltonian one can write down
is a two-band $\bm{k}\cdot\bm{p}$ like Hamiltonian (for valley +K) near $\bm{\kappa}'$ in the basis 
$(|j_t\rangle {\otimes} |\psi_{t,1,\bm{\kappa'}}\rangle, |j_b\rangle {\otimes} |\psi_{b,1,\bm{\kappa'}}\rangle)^T$,
where $|\psi_{t(b),1,\bm{\kappa'}}\rangle$ is the topmost moir\'e band state at $\bm{\kappa}'$ in each layer when $t=0$.
To the second order in momentum $\bm{q}$ (measured from $\bm{\kappa}'$), the effective Hamiltonian takes the form
\begin{align}
	\mathcal{H}_{\rm eff}(\bm{q}){=} 
	\begin{pmatrix}
		E_g/2+\hbar^2q^2/2m^{*}_t &  \hbar vqe^{i\Delta n\theta_{\bm{q}}} \\
		\hbar vqe^{-i\Delta n\theta_{\bm{q}}} & -E_g/2-\hbar^2q^2/2m^{*}_b
	\end{pmatrix}{,}
	\label{twobandHam}
\end{align}
where $\Delta n = j_t+m-j_b \mod\ 3$. 
$E_{g}$ is the bandgap and $\theta_{\bm{q}}$ is the orientation angle of $\bm{q}$. 
The effective parameters $m^*_t,\ m^*_b$, and $v$ can be obtained using perturbation theory in the non-interacting model (see details in the SM \cite{SM}).
In particular, the effective velocity $v$ is proportional to $t$.
$\Delta n\neq 0$ when the topological condition in Eq.~(\ref{topocond}) is satisfied.
When $\Delta n=0$, the lowest order off-diagonal term is a constant and $\mathcal{H}_{\rm eff}$ becomes trivial.

Within the two-band model, 
the transition from the NEI phase to the CI phase can be viewed 
as the switching of energetic order between $s$-wave and $p$-wave excitons upon increasing $t$.
It has been pointed out \cite{DiExciton, ImagoluBerry, YaoSciAdv} that in a gapped chiral fermion system,
the presence of the Berry curvature modifies the energy spectrum of excitons.
The energy change is proportional to the Berry curvature which is in turn proportional to $v^2$, $\Omega\propto v^2\propto t^2$,
as shown schematically in Fig.~\ref{fig:exciton}(b).
{\colorrev We further perform calculations on the interacting ground states of the effective model which well captures the phase competition (See details in the SM \cite{SM}).}

\emph{Discussions.}---\noindent
In many aspects the topology of the
moir\'e bands near the band inversion is similar to that
of the Kane-Mele model \cite{KaneMeleModel, WumoireTI, WangHF} or the BHZ model \cite{BHZmodel, FeiBHZ}.
However, it is only because of the narrow bandwidth, as a consequence of the moir\'e confinement,
that allows the ground state at $\nu=1$ to be fully valley polarized, exposing the bulk bandgap for the observation of QAH effect.
{\colorrev
The fate of the QAH state depends on the competition (or frustration) between Coulomb interaction 
and the pseudo-spin winding induced by the interlayer tunneling. 
This elementary mechanism has important implications for understanding the interplay between interaction and topology 
in strongly correlated topological moir\'e systems.
The valley polarized NEI state signifying this frustration
tends to appear in systems with weak interlayer tunneling and strong Coulomb interaction,
and can be probed 
by polarization-dependent optical spectroscopy \cite{ShanStripe} and angle-resolved transport measurement\cite{LeoAnisotropy},
both are demonstrated to be suitable for moir\'e systems.
(See detailed discussion in the SM \cite{SM}).
Moreover, the presence of the VPI$'$ phase suggests that mixing with remote moir\'e bands provides a possible path for
trivializing the CI phase without breaking spatial symmetry,
which can possibly explain the absence, so far, of the QAH effect in AA-stacked heterobilayers.
}

\begin{acknowledgments}
	{\em Acknowledgment.}---\noindent	The authors acknowledge helpful interactions with Zui Tao, Kin Fai Mak, Jie Shan, and Jiabin Yu.
	This work was supported by Laboratory for Physical Sciences.
	F. W. acknowledges support by National Key Research and Development Program of China (Grants No. 2021YFA1401300 and No. 2022YFA1402401), National Natural Science Foundation of China (Grant No. 12274333), and start-up funding of Wuhan University.
\end{acknowledgments}


\begin{thebibliography}{99}
\bibitem{withfield}
	With displacement field, the band inversion and interlayer hybridization
	only weakly modify the IVC phases 
	unless the topmost bands are deeply inverted.
\bibitem{gammatouching}
	We note that in the non-interacting model, 
	a subsequent band touching, after the band inversion, may occur at $\gamma$ between the first and second moir\'e bands,
	beyond which the Chern number vanishes.
	This does not affect the validity of the condition, Eq. (\ref{topocond}), near the band inversion.
\end{thebibliography}
\end{document}